\documentstyle[epsfig,psfig]{texas}

\begin{document}
\title{HARD X-RAYS FROM THE GALACTIC NUCLEUS: PRESENT AND FUTURE OBSERVATIONS}

\author{A.~Goldwurm $^{(1)}$, P.~Goldoni $^{(1)}$, P.~Laurent$^{(1)}$, 
F.~Lebrun$^{(1)}$, J.~Paul$^{(1)}$ }

\address{(1) Service d'Astrophysique/DAPNIA,  \\
CEA/Saclay, 91191 Gif sur Yvette Cedex, France \\
{\rm Email: agoldwurm@cea.fr}}

\begin{abstract}

In spite of increasing evidences of the presence of a massive Black Hole 
at the Galactic Center, its radio counterpart, Sgr~A$^{\rm *}$, shows little
activity at high energies, and recent models involving energy advection 
(ADAF) have been proposed to explain this difficulty.
We present results on the hard X-ray emission from the galactic central 
square degree obtained from the SIGMA/GRANAT 1990-1997 survey of this region.
The best upper limits available today on the Sgr~A$^{\rm *}$ 30-300~keV 
emission are presented and compared to X-ray data and to the 
predictions of ADAF models.
We also present simulations of Sgr~A$^{\rm *}$ observations with the future ESA
$\gamma$-ray Observatory INTEGRAL,
which show that the imager instrument (20~keV-10~MeV) will be able 
either to detect the expected ADAF emission from the accreting
Galactic Nucleus black hole 
or set tight upper limits which will constrain the physical parameters of 
such system.

\end{abstract}

\section{The Problem of the X/$\gamma$-Ray Low Luminosity of the Galactic 
Nucleus
}     

Increasing evidences in support of the presence of a massive 
(2.5~10$^{\rm 6}$~M$_{\odot}$) Black Hole (BH) at the dynamical center of 
our Galaxy have been collected in the past years. 
Proper and radial motions of stars in the central parsec of the Galaxy,
obtained with high resolution near-infrared observations \cite{ref1}, 
show indeed the presence of a dark mass with density 
$>$~10$^{\rm 12}$~M$_{\odot}$~pc$^{\rm -3}$ located within $<$~0.01~pc 
from Sgr~A$^{\rm *}$, and governing the dynamics of the mass of the region.

The compact synchrotron radiosource Sgr~A$^{\rm *}$, which
coincides with the Galaxy dynamical center and shows very low proper motion, 
is therefore considered
the visible counterpart of the Galactic Nucleus (GN) BH, and its spectrum has 
been studied at all wavelengths (for recent reviews on the Galactic Center see
\cite{ref2} \cite{ref3}). 
However, unlike stellar-mass BHs in binary 
systems and super massive BHs in AGNs, the Galactic Nucleus is found in the 
infrared and X-ray domains extremely faint and underluminous.
In particular the results of the Galactic Center SIGMA/GRANAT Survey 
\cite{ref4} \cite{ref5} \cite{ref6}
coupled to Rosat \cite{ref7}, ART-P \cite{ref8} and ASCA \cite{ref9} 
observations 
have shown that Sgr~A$^{\rm *}$ total X-ray luminosity is well below 
10$^{\rm 37}$~erg~s$^{\rm -1}$, i.e. $<$~10$^{\rm -7}$ times the Eddington 
Luminosity for such a BH.
This is rather intriguing since stellar winds of the close IRS~16 
star cluster provide enough matter to power the accreting BH.
Due to non uniformities in the winds the matter is accreted 
with substantial angular momentum and 
so the flow is certainly not in form of pure spherical free fall.
In standard thin accretion disks around BH about 10$\%$ of the energy 
provided by
the accreted matter ($\dot M$c$^{\rm 2}$) is expected to be radiated between 
infrared and X-ray frequencies. 
Estimates of accretion rate for Sgr~A$^{\rm *}$ are in the range
$(6-200)~10^{\rm -6}$~M$_{\odot}$~yr$^{\rm -1}$ \cite{ref10} \cite{ref11} 
and luminosities around $10^{\rm 40}-10^{\rm 42}$~erg~s$^{\rm -1 }$ 
are therefore expected.

New models of accretion flow have been recently proposed, in which very 
low radiation efficiency is obtained, even for non-spherical infall, by 
assuming 
that most of the energy is advected into the BH rather then being radiated. 
These ``advection dominated accretion flow'' models (ADAF) seem to be able to 
interpret the whole Sgr~A$^{\rm *}$ spectrum from radio to $\gamma$-rays, 
and in 
particular to explain the observed low X-ray flux \cite{ref12} \cite{ref11}
\cite{ref13}.
Sgr~A$^{\rm *}$ is in fact considered the test case for this set of models and 
observations in the X-ray domain of this object are crucial to establish the 
validity of the model assumptions and/or to constrain their parameters.

%--------------------------  figure 1
%this section shows how to insert a figure in the text
\begin{figure}
\centering
\psfig{file=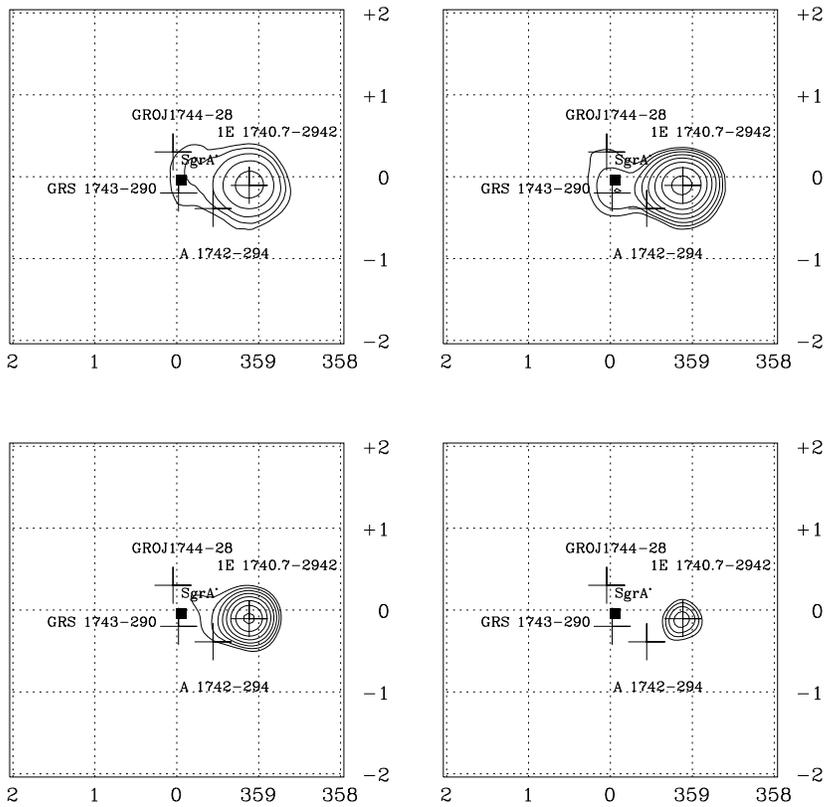, width=12cm}
\caption{
Images of the Galactic Center in galactic coordinates obtained from 
the 1990-1997 SIGMA Survey in 4 energy bands: 
30-40~keV up-left, 40-75~keV up-right, 75-150~keV bottom-left, 
150-300~keV bottom-right.
Contours are in units of standard deviations, starting from 4.5~$\sigma$ with
logarithmic steps of 1.4. Crosses indicate positions of SIGMA point sources
detected within the central degree in different periods \cite{ref6} 
while the black square shows the position of Sgr~A$^{\rm *}$.
}\label{fig1}
\end{figure}
%---------------------------------

\section{The 1990-1997 SIGMA/GRANAT Survey of the Galactic Center: 
30-300~keV Upper Limits on Sgr~A$^{\rm *}$}

The 30-1300~keV SIGMA telescope \cite{ref14} on the GRANAT satellite, observed
the Galactic Center between March 1990 and October 1997 about twice a year, 
for a total of 9.2~10$^{\rm 6}$~s effective time.
The telescope provided an unprecedented angular resolution ($15'$) at these 
energies and, for the quoted observing time, a
typical 1$\sigma$ flux error of $<$ 2-3 mCrab (1 mCrab is about 8 $\times$
10$^{-12}$ erg cm$^{-2}$ s$^{-1}$ in the 40-80 keV band).
Analysis of a subset of these data 
already provided the most precise hard X-ray images of the Galactic Nucleus
\cite{ref4} \cite{ref5}
and proved that the Sgr~A$^{\rm *}$ luminosity in the 40-150 keV band is 
lower then 10$^{\rm 36}$~erg~s$^{\rm -1}$.

These results were important because, though it was known from the Einstein 
Observatory data \cite{ref15}, that the GN was not bright in soft X-rays,
it was still possible that Sgr~A$^{\rm *}$ could have,
as any respectable BH in low state, a bright hard tail extending 
to $>$~100 keV and possibly also a component of 511~keV line emission. 
For example the close source 1E~1740.7-2942, weak at $<$~4~keV was later 
observed, in particular by SIGMA,
to be bright in soft $\gamma$-rays, was found associated to 
radio jets and even to display transient events of 500~keV emission (see 
references in \cite{ref4}).
Indeed 1E~1740.7-2942 is now recognized to be a good BH candidate at only 
40$'$ from the GN and to be responsible for most of the
hard X-ray/$\gamma$-ray activity observed by early 
low-angular-resolution experiments from the direction of the Galactic 
Center and previously associated to the GN \cite{ref16}.

Recently Goldoni et al. (1999) \cite{ref6}, have re-analyzed the full set of 
SIGMA data to improve the upper limits estimation. 
In this analysis they included models 
with both point-source and diffuse emission to fit sky images. Reconstructed 
images of the central 4$^\circ$$\times$4$^\circ$ region in different energy 
bands are presented in Fig.~1. 
Four variable point-sources are contributing to the emission from the central 
1$^\circ$ circle (1E~1740.7-2942, A~1742-294, GROJ~1744-28, GRS~1743-290). 
The last one is only $11'$ from Sgr~A$^{\rm *}$ and, 
if considered a single source, cannot be associated to Sgr~A$^{\rm *}$. 
However the presence of some weak contribution 
from the Galactic Nucleus cannot be excluded and therefore 2 sets of limits
were derived for the Sgr~A$^{\rm *}$ hard X-ray flux: 
one for which all GRS~1743-290 emission is attributed to one single source
(case A) and another obtained by including a source at the Sgr~A$^{\rm *}$ 
position 
(fixed parameter) in the fitting procedure (case B). Results are reported in 
Table~1, for the four energy bands of Fig.~1. At high energy no emission at 
all is detected apart from 1E~1740.7-2942 and values in the two colums are 
identical.

\begin{table}
\begin{center}
\begin{tabular}{ccc}

\multicolumn{3}{c}{}\\
\multicolumn{3}{c}{\large \bf SIGMA/GRANAT Upper Limits on Sgr~A$^{\rm *}$}\\
\multicolumn{3}{c}{}\\

\hline
\noalign{\smallskip}
{\rm Energy~range}&{\rm Integrated~luminosity~(A)}&{\rm Integrated~luminosity~(B)}\\
{\rm keV} & {\rm erg~s$^{\rm -1}$} & {\rm erg~s$^{\rm -1}$} \\
\noalign{\smallskip}
\hline
\noalign{\smallskip}
$30-40~{\rm keV}$  & $2.6 \times 10^{35}$ & $4.5 \times 10^{35}$\\
\hline
\noalign{\smallskip}
$40-75~{\rm keV}$  & $2.0 \times 10^{35}$ & $3.4 \times 10^{35}$\\
\hline
\noalign{\smallskip}
$75-150~{\rm keV}$  & $2.0 \times 10^{35}$ & $2.4 \times 10^{35}$\\
\hline
\noalign{\smallskip}
$150-300~{\rm keV}$  & $5.2 \times 10^{35}$ & $5.2 \times 10^{35}$\\
\hline
\noalign{\smallskip}

\end{tabular}
\vspace{3mm}
\caption{SIGMA/GRANAT 2$\sigma$ upper limits for Sgr~A$^{\rm *}$ luminosity}
\label{Table1}
\end{center}
\end{table}

The most stringent upper limits of Table~1 (A column) have been reported 
in Fig.~2 in units of E~L$_{\rm E }$ for a distance of 8.5 kpc, and
compared with the ADAF model predicted spectrum of Sgr~A$^{\rm *}$. 
The reported model is the ``best model'' as defined in \cite{ref11}, and 
refers to a BH mass of 2.5~10$^{\rm 6}$~M$_{\odot}$, 
a mass accretion in Eddington units ${\dot m}$~=~1.3~10$^{\rm -4}$ 
(i.e. 7~10$^{\rm -6}$~M$_{\odot}$~yr$^{\rm -1}$), 
a viscosity parameter $\alpha$~=~0.3,
an equipartition parameter $\beta$~=~0.5 
(exact equipartition between gas pressure and magnetic pressure), 
and a fraction of viscous heat converted in electron heat of 
$\delta$~=~0.001.

Comparison with X-ray results, i.e. 
luminosities measured by Rosat (0.8-2.5~keV) and
ASCA (2-10~keV), is also shown in Fig.~2. ASCA value is actually reported as 
upper limit, following Narayan et al. 1998 \cite{ref11}, because the observed 
flux was rather associated to the X-ray burster A~1742-289, located only 
at 1$'$ from Sgr~A$^{\rm *}$ \cite{ref9}. 
Note however that this interpretation is controversial and the ASCA measure 
may well contain some contribution from Sgr~A$^{\rm *}$ \cite{ref17}. 

%--------------------------  figure 2
%this section shows how to insert a figure in the text
\begin{figure}
\centering
\psfig{file=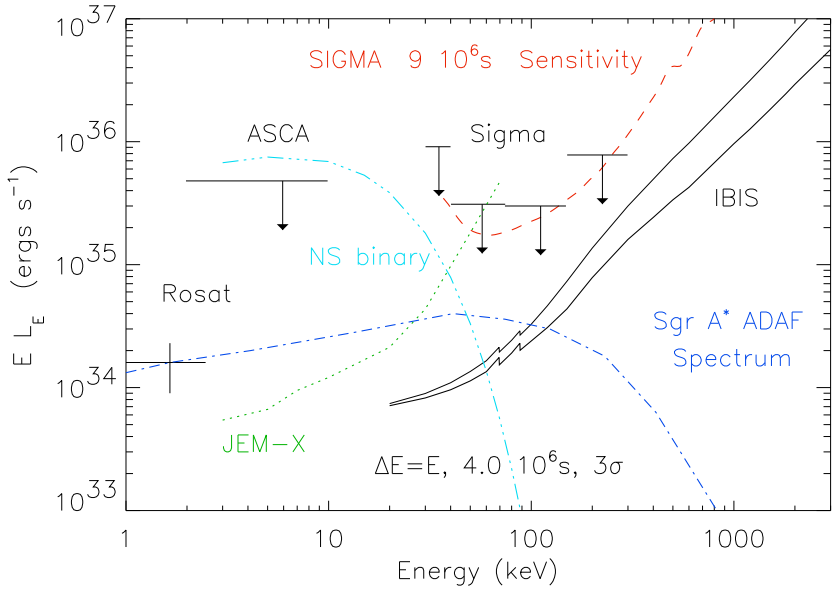, width=12cm}
\caption{
Sgr~A$^{\rm *}$ 30-300~keV upper limits in units of E~L$_{\rm E }$ (d=8.5 kpc),
obtained from the 1990-1997 SIGMA/GRANAT Galactic Center Survey, compared
to the predicted ADAF spectrum (for M$_{BH}$=2.5~10$^{\rm 6}$~M$_{\odot}$ 
and ${\dot m}$=1.3 10$^{\rm -4}$) \cite{ref11} (blue dashed-dotted line), 
along with the SIGMA sensitivity (red dashed line) scaled to the survey time
of 9~10$^{\rm 6}$~s. Rosat Sgr~A$^{\rm *}$ flux and the ASCA upper limit are 
also reported.
IBIS (black full lines) and JEM-X (green dotted line) broad-band sensitivities
for the Galactic Nucleus and the INTEGRAL Galactic Center Deep Exposure time 
of 4~10$^{\rm 6}$~s are also reported
(JEM-X  sensitivity at 20~keV is lower then IBIS one due to smaller FOV).
A typical thermal spectrum (kT$\approx$10~keV) of a neutron star LMXB
(1E~1743.1-2864, \cite{ref8}) 
(light-blue dashed-double-dotted line) is also shown for comparison.
}\label{fig2}
\end{figure}
%---------------------------------

\section{The IBIS/INTEGRAL Galactic Center Deep Survey}

The Imager on Board the Integral Satellite (IBIS) is one of the two main
instruments of INTEGRAL, the ESA $\gamma$-ray mission to be launched in 2001.
It provides, thanks to its coded-aperture imaging system composed by a  
tungsten mask and two pixellated detector layers,
fine imaging (12$'$ FWHM resolution), good spectral resolution ($<$~8~$\%$ 
at 100~keV and $\approx$~6~$\%$ at 1~MeV) and good sensitivity over a wide
(20~keV-10~MeV) energy range  and a very wide field of view 
(29$^\circ$$\times$29$^\circ$ at 0 sensitivity) \cite{ref18}. 
The INTEGRAL Core Program includes the so called Galactic Center Deep Exposure
(GCDE) program, the deep observation of a central Milky Way region 60$^\circ$ 
wide in longitude and 20$^\circ$ in latitude. 
The GCDE will consist of a grid of 31$ \times $11 pointings separated by 
2$^\circ$.
The net GCDE observing time is 4.8 10$^{\rm 6}$~s per year, of which about 
84~$\%$ performed in pointing mode \cite{ref19}.
Considering the GCDE scan in pointing mode and the IBIS sensitivity over 
its FOV we estimated that the Galactic Nucleus will be actually observed 
by IBIS for an effective (on-axis equivalent) time of 
$\approx$~4~10$^{\rm 6}$~s in 3 years (see Fig.~3 for the IBIS/Sensitivity
of the INTEGRAL GCDE scan).
Adding slew and Galactic Plane Survey times the total net
3-years IBIS Core Program exposure on the GN will increase 
to 5~10$^{\rm 6}$~s. 
IBIS broad-band sensitivity is shown in Fig.~2 for the pointing GCDE time 
on the E~L$_{\rm E}$~vs.~E 
plot to compare it with the ADAF model spectrum for Sgr~A$^{\rm *}$ 
\cite{ref11}. Two curves for IBIS are represented for two extreme 
estimates of the in-flight background.
In Fig.~2 is reported also the expected GCDE sensitivity
for the INTEGRAL X-Ray monitor (JEM-X) which provides images in the range
3-60~keV but in a smaller field of view. 
This picture demonstrates that IBIS sensitivity estimated for the GCDE will 
allow either to detect the high energy emission predicted by ADAF
from Sgr~A$^{\rm *}$ or to set tighter constraints on the model parameters.

%--------------------------  figure 3
%this section shows how to insert a figure in the text
\begin{figure}
\centering
\psfig{file=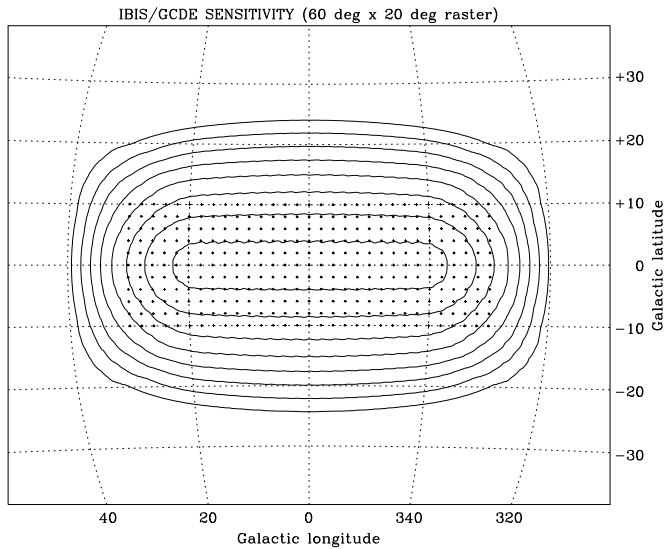, width=9cm}
\caption{
IBIS Sensitivity for the INTEGRAL CGDE Survey, as fraction of on-axis
IBIS sensitivity for the same exposure time. Contours are 
between 0.06 and 4.8 with steps of 0.06. Crosses indicate pointing
scan of the GCDE \cite{ref19}. Maximum value (at l=0$^{\circ}$ b=0$^{\circ}$)
is 0.51.
}\label{fig3}
\end{figure}
%---------------------------------

\section{ Simulations of the Galactic Nucleus IBIS/INTEGRAL Observations}

ASCA, ART-P and Rosat results have shown that at low energies lots of point 
sources and also diffuse 
emission are present in the area and can make identifications difficult. 
In particular activity
of the X-ray burster A1742-289, only 1$'$ away from Sgr~A$^{\rm *}$, 
would not be easily separated by the GN one even by X-ray instruments
like JEM-X.
On the other hands results at higher energies, i.e. around 80~keV, should be
rather ideal to reveal emission from the GN, since at these energies, 
due to their softer spectra, most of the Neutron Star binaries will be 
significantly fainter than the expected emission from the GN BH 
(see NS binary typical spectrum in Fig.~2).
Even at energies $>$~50~keV, however, imaging will be crucial since SIGMA 
showed the presence of several high-energy sources located within 2$^\circ$
from the Galactic Nucleus (see Fig.~1).

To prove the imaging capabilities of the IBIS telescope we performed 
simulations of a deep IBIS observation of the Galactic Center with the ISGRI 
low energy (20-700 keV) $\gamma$-ray detector layer of the telescope 
\cite{ref20}.
Basic simulation procedure was described in \cite{ref21} \cite{ref22},
we used results of the SIGMA survey for source fluxes \cite{ref4} and the 
predicted Sgr~A$^{\rm *}$ flux from ADAF spectrum.
Sky image reconstruction procedures are based on standard cross-correlation 
techniques and iterative analysis and removal of sources as described e.g. 
in \cite{ref23}.
One simplified assumption was 
that the observation was performed as a single pointing rather
then a sum of pointings along a grid as it is actually expected. 
This should not influence the results since the scanning will rather
help to remove background unknown systematic effects, 
not presently included in the simulation.
Work is in progress to include more realistic operational conditions.

Fig.~4 shows some of the results of the simulations. 
These are zooms of the central part of reconstructed sky images
obtained by an iterative decoding and point-source cleaning algorithm applied 
to the simulated detector images. 
A cluster of 4 high energy sources in the central 2$^\circ$ circle appears in 
the images (Fig.~4, left). Fine image analysis allow to position them and 
to remove their contribution (Fig.~4, right).
Sgr~A$^{\rm *}$ is then detected at the expected position and signal to noise 
ratio of $\approx$~6~$\sigma$, corresponding to 
L$_{\rm 50-140~keV}$~$\approx$~3.5~10$^{34}$~erg~s$^{-1}$, the level predicted
by the ADAF model.
Including diffuse emission (possibly present at $<50$~keV energies
\cite{ref9}) showed that its presence can make detection of faint sources more 
complicated and somemore refined procedures should be employed to model
the composite diffuse plus point-sources emission.
However diffuse emission will not influence the search for the GN emission
at high energies.

%--------------------------  figure 4
%this section shows how to insert a figure in the text
\begin{figure}
\centering
\psfig{figure=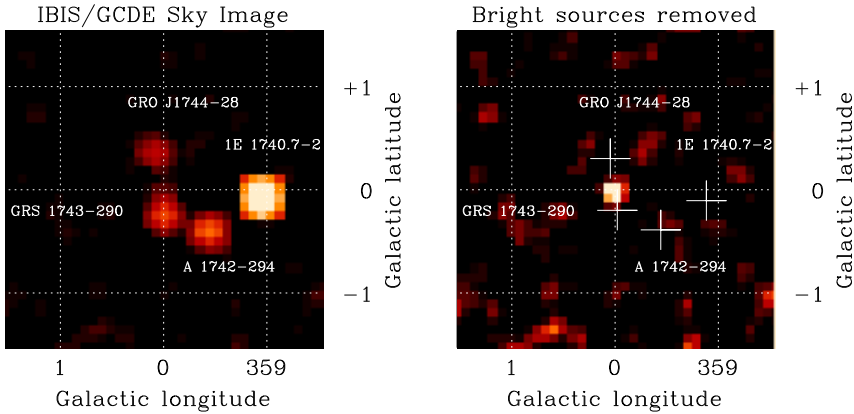,width=12cm}
\caption{
The central 3$^\circ$$\times$3$^\circ$ of the deconvolved and 
cleaned sky images in standard deviations (1 pixel $\sim~5'$), 
from a simulation of the 
IBIS/ISGRI GCDE observation (4~10$^{\rm 6}$~s) in the 
50-140 keV band (logarithm scale from 1 to 200~$\sigma$ left, 
linear scale from 1 to 5~$\sigma$ right). 
Simulation included 11 sources in the IBIS FOV with fluxes from 
4 to 70 mCrabs \cite{ref4}, a background of 150~cts~s$^{\rm -1}$ 
resulting in 1~$\sigma$ imaging error of 0.11 mCrab and a 0.6~mCrab
point-source in Sgr~A$^{\rm *}$ (L$_{50-140~keV}$ 
$\approx$~3.5~10$^{34}$~erg~s$^{-1}$). 
After removal of the 4 brightest 
sources Sgr~A$^{\rm *}$ is detected at the simulated
position and expected S/N level of $\approx$~6~$\sigma$.
}
\label{fig4}
\end{figure}

\section{Conclusions: XMM Core Program Observations of Sgr~A$^{\rm *}$}

We have presented here the best available hard X-ray upper limits on 
Sgr~A$^{\rm *}$, 
obtained by the deep Galactic Center SIGMA/GRANAT survey. As shown above 
(Fig.~2), they do not constrain the present ADAF models, 
invoked to resolve the apparent contradiction between the presence
of a massive black hole at the Galactic Center and its lack of activity in the
X-ray domain. However we note that the ADAF model critically depends on
mass accretion since L~$\propto$~${\dot m}^{\rm 2}$.
The assumed value of accretion rate in the model 
(${\dot m}$=$1.3~10^{\rm -4}$ \cite{ref11}, 
actually determined by Rosat flux) is close
to the lower limit of the estimated range of ${\dot m}$ 
(i.e. $(1-30)~10^{\rm -4}$, \cite{ref11} \cite{ref10}). 
Even a factor 3 higher in ${\dot m}$ would make the model not consistent 
with our limits,
and also not compatible with other spectral data of Sgr~A$^{\rm *}$.

New generation of telescopes aboard the future X-ray (AXAF/Chandra, XMM) 
and gamma-ray (INTEGRAL) missions, will allow to deeply search and study 
the GN high energy emission and in this way to test present models 
for massive BH accretion.

We have shown with simulations that IBIS/INTEGRAL will be able to disentangle 
the hard X-ray emission of the Milky Way central square degree and to detect
ADAF emission from Sgr~A$^{\rm *}$ or to set appropriate upper limits over 
the band 50-140~keV, by using 3 year data of the INTEGRAL/GCDE core program.
In case of detection, the reconstructed flux will allow to test ADAF spectra 
and the radiation processes expected in such a massive BH.
On the other hands the set of upper limits will imply an $\dot M$
at least a factor 1.4 lower then the present value of ADAF model, making 
the problem of low accretion rate even more difficult.
The ADIOS models \cite{ref24}, in which advection is coupled 
to inflows/outflows of matter and which allow even lower 
emission then ADAF models for the same mass supply rate, may have then to be
invoked.
Alternatively different hypothesis on magnetic field will have to be included,
e.g. allowing for lower values of the $\beta$ equipartition parameter in order 
to steepen the spectrum and reduce contribution at high energies \cite{ref11}.

In this context hard X-ray results will be even more valuable if coupled 
to high resolution results at lower energies which could constrain the mass 
accretion rate and allow to study the spectral shape. 
XMM observations of the Galactic Center are planned in the Core Program 
of the first year of the mission operations. More than 5~10$^{\rm 4}$~s 
of the XMM
Galactic Center scan will be devoted to observe the central half-degree 
of the Galaxy with the EPIC cameras.
Such exposure time will provide sensitivities 
of the order of 10$^{32}$~erg~s$^{-1}$ ($5~\sigma$ for 8.5 kpc distance) 
in the energy range $2-10$~keV, 
with 6$''$ angular resolution (FWHM) and good spectral resolution.
We expect that these observations will allow to resolve the confusion over 
the soft X-ray sources associated to Sgr~A$^{\rm *}$ \cite{ref17} and 
to obtain spectral data to test radiative properties of the closest
supermassive accreting black hole.

\end{document}